\documentclass[conference]{IEEEtran}
\IEEEoverridecommandlockouts
\usepackage{cite}
\usepackage{amsmath,amssymb,amsfonts}
\usepackage{algorithmic}
\usepackage{graphicx}
\usepackage[braket, qm]{qcircuit}

\usepackage{caption}
\usepackage{subcaption}
\usepackage{multirow}
\usepackage{textcomp}
\usepackage{xcolor}
\def\BibTeX{{\rm B\kern-.05em{\sc i\kern-.025em b}\kern-.08em
    T\kern-.1667em\lower.7ex\hbox{E}\kern-.125emX}}
\begin{document}

\title{AI methods for approximate compiling of unitaries} 

\author{
\IEEEauthorblockN{
David Kremer, 
Victor Villar, 
Sanjay Vishwakarma,
Ismael Faro 
and Juan Cruz-Benito}
\IEEEauthorblockA{
IBM Quantum. IBM T.J. Watson Research Center, \\Yorktown Heights, NY, USA}}

\maketitle

\begin{abstract}
This paper explores artificial intelligence (AI) methods for the approximate compiling of unitaries, focusing on the use of fixed two-qubit gates and arbitrary single-qubit rotations typical in superconducting hardware. Our approach involves three main stages: identifying an initial template that approximates the target unitary, predicting initial parameters for this template, and refining these parameters to maximize the fidelity of the circuit. We propose AI-driven approaches for the first two stages, with a deep learning model that suggests initial templates and an autoencoder-like model that suggests parameter values, which are refined through gradient descent to achieve the desired fidelity. We demonstrate the method on 2 and 3-qubit unitaries, showcasing promising improvements over exhaustive search and random parameter initialization. The results highlight the potential of AI to enhance the transpiling process, supporting more efficient quantum computations on current and future quantum hardware.
\end{abstract}


\section{Introduction}

Transpiling is a critical stage in the current quantum computing workflows, and consists of adapting a given circuit to a set of target instructions supported on specific quantum devices. Modern quantum circuit transpilers include several passes dedicated to different tasks \cite{Qiskit,javadi2024quantum,2405.13196}, and aim to adapt the given quantum circuit to the hardware with as little gate and depth overhead as possible to reduce the effects of noise on the quantum hardware.


Unitary Synthesis is a fundamental task within most transpilers. It consists of, given a unitary matrix, finding a sequence of gates within the hardware's instruction set that implement the given matrix.

In this paper we focus on exploring how AI methods can improve Unitary Sythesis, a central part within typical transpiling workflows. We focus on the case where the instruction set is composed of fixed two-qubit gates (such as CZ or CX) and single qubit rotations (such as RZ) that can implement arbitrary rotations, as is common in current superconducting hardware \cite{chow2012universal,niskanen2007quantum}. 

Within this setting, exising algorithms broadly fall into two categories:

\begin{itemize}
    \item Exact methods, such as the two qubit KAK-based decomposition described in \cite{Cross_2019} or the more general Quantum Shannon Decomposition  \cite{Drury_2008, Shende_2006} that exactly implement the given unitary. These methods provide an upper bound on the number of two-qubit gates for general unitaries, but except for the two qubit case, the circuits they produce are far from optimal for unitaries that can be found in typical applications. 
    \item Approximate methods that, based on a given circuit template (a circuit with fixed gates and adjustable single qubit rotations), adjust the circuit parameters to maximize the fidelity of the circuit with respect to the given unitary \cite{robertson2023approximate, madden2022best}.
\end{itemize}

Although the approximate methods scale beyond 2 qubits, they have some drawbacks that make their widespread use impractical for transpiling:
\begin{itemize}
    \item Template selection is not straightforward, and most methods resort to search methods that require iterating over several templates to find one that can successfully implement the target unitary (up to some high fidelity threshold).

    \item Optimization of the parameters is time consuming, and it often reaches local minima, resulting in sub-optimal fidelities even when the given template is capable of reaching perfect fidelity for the target unitary.
\end{itemize}

There have been recent attempts to improve this process. In \cite{robertson2023approximate}, they use tensor networks to reduce the amount of computation during the optimization, allowing them demonstrate approximate compiling on shallow circuits on 27 qubits, but only for state preparation and not for the full unitary transformation. 

In \cite{weiden2023improving}, they propose a tree-based search method where they train a deep learning model to predict which template to use from a set of 3-qubit templates, and start a tree-based template search from there, but they don't address the problem of parameter optimization.

Here we explore a variety of Machine Learning based methods to potentially speed up the process and allow synthesis of larger unitaries on typical transpiling pipelines, and propose simple AI-based algorithms for template selection and circuit parameter optimization.

The paper is organized as follows: Section II describes the general approach and the Machine Learning methods used and the data generation procedures used for the template selection and the parameter optimization. Section III shows the results obtained by the method for 2 and 3 qubit unitary synthesis. Finally, in Section IV we discuss the results and future research directions of interest.





\section{Materials and methods}

For addressing the unitary synthesis problem, we develop a pipeline for unitary synthesis composed of three stages:

\begin{enumerate}
    \item \textbf{Template selection.} A deep learning model that suggests a template given a target unitary.
    \item \textbf{Parameter prediction.} A deep learning model that suggests an initial value for the template's parameters as a starting point for the optimization.
    \item \textbf{Parameter optimization.} Further Optimization of the parameters via gradient descent, up to the desired fidelity.
\end{enumerate}

For synthesizing a given unitary, the unitary is first assessed by the template selection model, which suggests a first template, and then the unitary is passed to the second model that provides suggested values for the template. These values are then used as starting point for an optimization via gradient descent. If the desired fidelity is not reached, the procedure starts again by selecting the second template suggested by the model, and continues until the objective fidelity is reached, or defaults to other methods if the suggestions are exhausted. 

When used in the context of circuit optimization, where circuit blocks are collected from a larger circuit and re-synthesized, the two qubit depth and count of the block presents an upper bound on the template size. This can be used to filter the suggestions to the ones that incur lower cost. If the model is confident that the cost of a block cannot be improved, the block can be left as is without attempting optimization.

For the cases when the template selection model guesses correctly, and where the second model suggests a useful starting point, this may result in significant speedups with respect to exhaustive template search, and might result in better fidelities by avoiding local minima.

In the next subsections we focus on the first and second stages and describe the network architectures used and the data generation processes.





\subsection{Template prediction}

For the first stage we use supervised deep learning techniques to select the right template to be used in the second step. 

\subsubsection{Network architecture and training}

We train a multi layer neural network to predict a template, which poses a supervised classification problem. The input to the neural network is a vector with the raw values of the entries of the unitary matrix, split in their real and imaginary parts (this gives a total number of elements of $2 \cdot 2^{N}$, being N the number of qubits). For our experiments, we have seen that simply using standard fully connected layers for the network architecture is enough for achieving high accuracy.

The output of the network is a vector with the probabilities assigned to each of the possible templates. Since we only allow one "correct" template for a given unitary, we train the network to minimize the cross entropy loss between the probabilities provided by the network and the template for which the unitary was generated. As it is often done in multiclass classification, the last layer of the neural network normalizes the output with a softmax so that the probabilities sum up to one.

\subsubsection{Data generation}

During the training, we generate batches of unitary-template pairs on the fly. This ensures an "unlimited" supply of data and acts as a form of regularization preventing overfitting to a particular fixed set of unitaries. 

\begin{figure}[htbp]
\centering
\begin{subfigure}{.4\textwidth}
\centering
\includegraphics[width=1.0\linewidth]{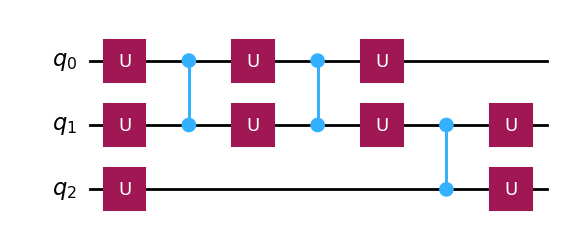}
\caption{Template 9}
\end{subfigure}
\begin{subfigure}{.4\textwidth}
\centering
\includegraphics[width=1.0\linewidth]{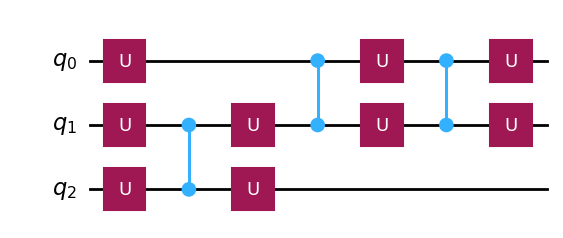}
\caption{Template 12}
\end{subfigure}
\caption{Examples of two templates for 3 qubits.}
\label{templates_3q}
\end{figure}

We first define a set of templates for which we want to predict. Note that this doesn't need to include the complete set of possible templates for a given number of qubits; in particular we are typically interested in shallower unitaries that appear frequently on larger circuits, and we can always include an extra term in the model's output to indicate if a given unitary is not implementable in any of the templates of the set. In our experiments, we build the templates by combining single-qubit unitary gates and two-qubit CZ gates. A few examples of templates for 3 qubits can be found in Figure \ref{templates_3q}.

For generating each row of the batch, we first randomly pick a template from the template set, which will be the label that will be predicted by the model. We then generate a random instance of the chosen template by assigning random values for the parameters of the single-qubit unitaries, and we obtain the unitary matrix corresponding to the circuit. 

\begin{figure*}[htbp]
\centering
\begin{subfigure}{.5\textwidth}
\centering
\includegraphics[width=1.0\linewidth]{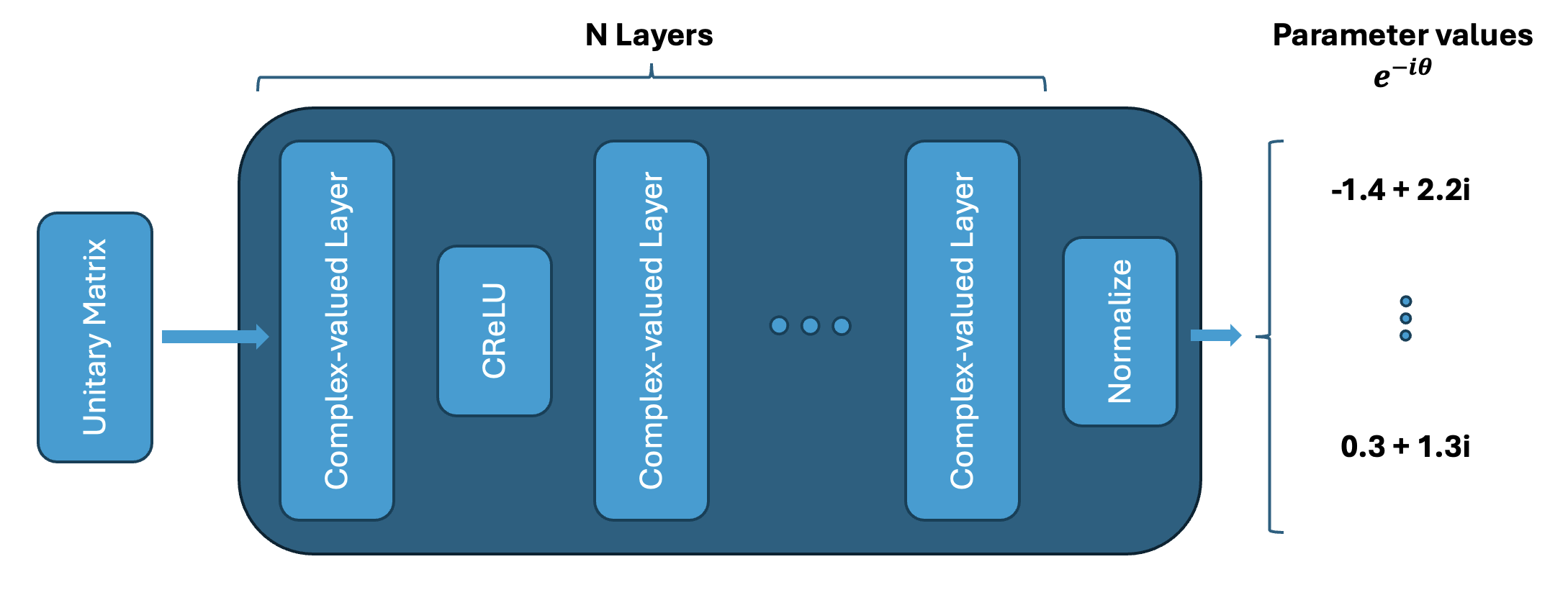}
\caption{Encoder network}
\label{encoder}
\end{subfigure}%
\begin{subfigure}{.5\textwidth}
\centering
\includegraphics[width=1.0\linewidth]{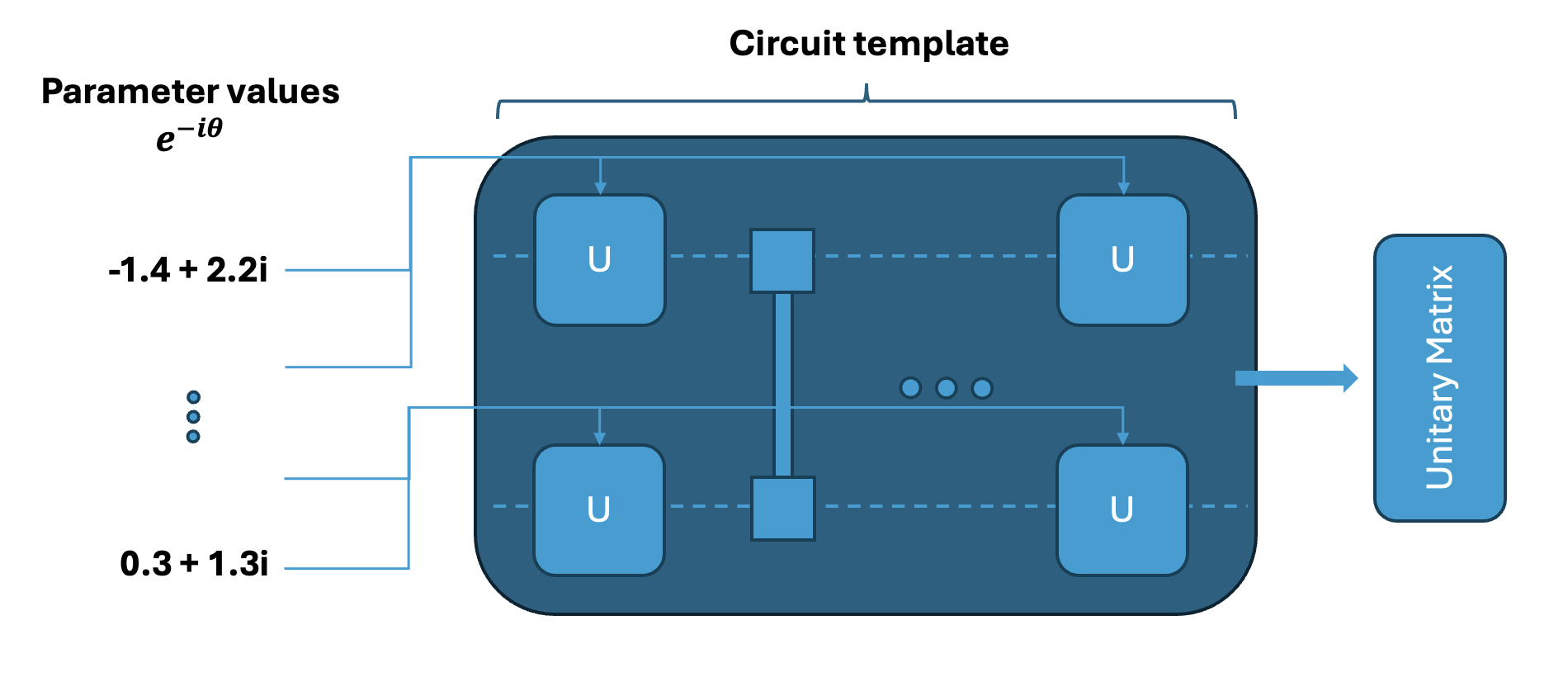}
\caption{Decoder}
\label{decoder}
\end{subfigure}
\caption{Encoder and decoder networks for template parameter suggestion}
\label{encoder_decoder}
\end{figure*}


\subsection{Parameter suggestion}

For the second stage we define a network architecture and training procedure to generate good suggested values for the template parameters for a given template. 

A naive approach for predicting good parameters for a given unitary would be to simply train a neural network model on unitary-parameter pairs. For generating a unitary-parameter pair one would first uniformly sample a group of parameters from the allowed parameter range (typically $[-\pi,\pi]$ for rotation angles), and then obtain the corresponding unitary by setting these parameters in the template. Repeating this for multiple parameter values would yield a dataset for training. However, we have found that with this method the models fail to produce useful results, or to learn anything at all. Because a given unitary can be implemented by multiple (very different) sets of parameters for the given template, by using this procedure we may generate multiple different target parameters for unitaries that are potentially very close. This results in a dataset of unitary-parameters pairs that cannot be described by a mapping that assigns a single value for the parameters given a unitary matrix, and prevents the network from learning any useful relationship.

Although one could overcome this by generating the dataset in a consistent manner, for example by first sampling a unitary matrix and then using gradient descent from a fixed starting point, this approach still has some fundamental limitations. One of the main drawbacks is that by training on unitary-parameter pairs we are teaching the network to, in some way, interpolate between the parameters in the dataset; however, the actual objective is not to provide parameters that mimick the dataset, but to provide parameters that maximize the fidelity of the circuit.

\subsubsection{Network architecture and training}

For overcoming this limitation, we define a deep learning model based on the autoencoder architecture \cite{hinton2006reducing, bengio2009learning,bank2023autoencoders}, consisting on an encoder network and a decoder. Here, the encoder is a deep learning network (as in standard decoders) that takes the unitary matrix as input and outputs a value for each of the parameters on the template. The decoder, unlike typical autoencoders, is fixed, and corresponds to the simulation of the actual template of the quantum circuit from the parameters output by the encoder. 

For training, we first take an input unitary and pass it through the encoder to generate the predicted parameters. We then pass this parameters through the decoder to reconstruct the unitary. With both unitaries, we calculate the reconstruction fidelity of the output unitary with respect to the input unitary. Since the simulation of the circuit is also differentiable, we backpropagate the gradients of the fidelity with respect to the encoder network's weights and use them to train the network. This allows us to train the network directly to maximize the reconstruction fidelity.

For the encoder network architecture, we use again fully connected layers, but this time we choose complex-valued weights for the trainable parameters. This allows us to directly input the complex values of the unitary matrix and output complex values for the entries of the single-qubit unitary matrices of the template that represent the template's parameters.

An illustration of the encoder and decoder architecture can be found in Figure \ref{encoder_decoder}.

\subsubsection{Data generation}

With the autoencoder architecture, our dataset now only consists of "x" points that correspond to sampled unitaries, since we don't need to generate any parameters to train on. For the training, as in previous sections, we generate the unitaries on the fly. We generate the unitaries by sampling from the given template, to ensure they can be implemented on that template. 

\section{Results}

\begin{figure*}[htp]
\centering
\begin{subfigure}{.5\textwidth}
\centering
\includegraphics[width=0.9\textwidth]{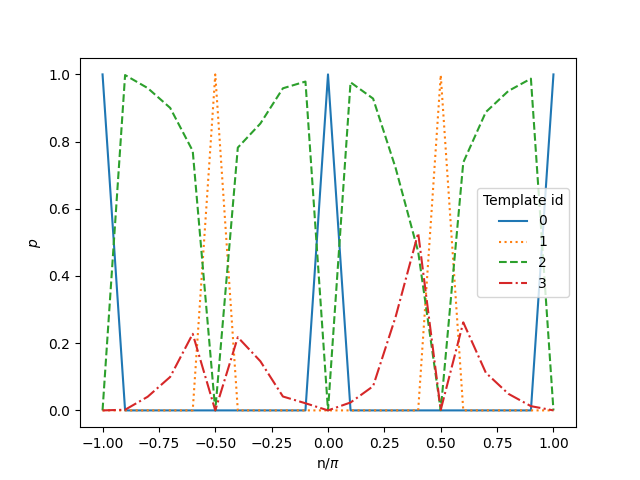}
\caption{Probability assigned to each template as a function of $n/pi$.}
\label{template_pred_probs}
\end{subfigure}%
\begin{subfigure}{.5\textwidth}
\centering
\includegraphics[width=0.9\textwidth]{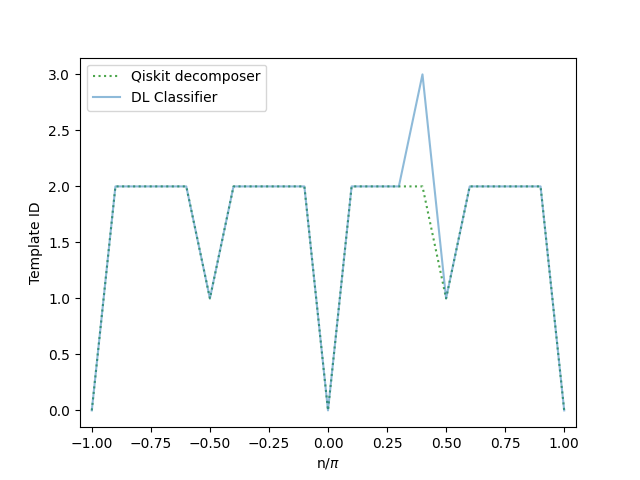}
\caption{Template predicted as a function of $n/pi$.}
\label{template_pred_vs_qiskit}
\end{subfigure}
\caption{Probability assigned to each template and predicted template as a function of $n/pi$, where $n$ is the angle used in gate CZ for the circuit described in the text.}
\end{figure*}

In this section we show the results obtained in training models for 2 and 3 qubits for both stages. For the models we trained, we selected the instruction set of the IBM Quantum Heron devices \cite{gambetta2022quantum}: we use CZ for the two-qubit gates, and implement the single-qubit unitaries as a sequence of 3 CZ rotations separated by SX gates.

\subsection{Template prediction}
For the 2 qubits model, we use 4 different templates, containing from 0 to 3 CZ gates respectively. As pointed out in \cite{Cross_2019}, this set of templates is complete in the sense that it allows to implement any two qubit unitary with optimal number of two qubit gates. 

For 3 qubits, we train two models: one for 15 different templates, covering all possible combinations using up to 3 CZ gates, and another for 63 possible templates, covering all possible combinations up to 5 CZ gates. This set of templates is not enough to cover all possible unitary circuits, but it is useful for optimizing small blocks that typically appear on circuits. 

In order to assess if model capacity was a limiting factor, we train multiple models of different sizes. The results are summarized on Table \ref{table_template_parameters}.

\begin{table}[htbp]
\caption{Template prediction training parameters.}
\begin{center}
\begin{tabular}{|c|c|c|c|}
\hline
&\multicolumn{2}{|c|}{\textbf{\textit{Train configuration}}}& \textbf{\textit{Validation result}}\\
\hline
\textbf{\textit{\#Qubits}}&\textbf{\textit{\#Network Parameters ($10^6$)}} & \textbf{\textit{Size (MB)}}& \textbf{\textit{Accuracy}}\\
\hline
\multirow{3}{*}{2}& 0.2 & 0.8 & 0.95 \\
& 7.4 & 29.5 & 0.96 \\
& 12.6 & 50.6 & 0.97 \\
\hline
\multirow{3}{*}{3  (3 CZ)}& 3.3 & 13.2 & 0.74 \\
& 5.4 & 21.5 & 0.75 \\
& 21.3 & 85.1 & 0.77 \\
\hline
3  (5 CZ) & 9.6 & 38.5 & 0.37 \\
\hline
\end{tabular}
\label{table_template_parameters}
\end{center}
\end{table}

For 2 qubits we reach a validation accuracy of $\simeq96\%$, with a difference of only $2\%$ between the worst and the best models. Even if we see a trend on decreasing accuracy with decreasing model sizes, the difference is small and may be improved with different training hyperparameters or longer training time. Overall, an interesting thing to note is that we can reach high accuracy with very small models, with the smallest one we tested being only $0.8$ MB.

To test the robustness of the template prediction for 2 qubits, we run a simple experiment where we predict the template for a circuit consisting of a two consecutive CNOT gates separated by an RZ gate on the first qubit. We vary the angle of the RZ gate from $-\pi$ to $\pi$, allowing us to test how the model transitions between templates: the two-CZ template for most of the range, the zero-CZ template when the angle is a multiple of $\pi$ (allowing the two CNOTs to cancel out) and a less evident one-CZ template when the angle is $\pm\pi/2$.

In Figure \ref{template_pred_probs} we can see the probability the model assigns to each template (with 0, 1, 2 or 3 CZ gates). The model predicts with high probability the right template in most of the range, except around $n/\pi\simeq0.45$, where it incorrectly predicts the template with 3 CZ gates instead of the one with 2 CZ gates by a close margin. In figure \ref{template_pred_vs_qiskit} we validate the predictions obtained against the solution provided by the $Qiskit$ transpiler (using the KAK-based method described in \cite{Cross_2019}).

For 3 qubits with up to 3 CZ, the best model reaches an accuracy of $\simeq77\%$ when selecting the highest probability template. If we also consider the second highest probability, the correct template is present in this set approximately $\simeq98\%$ of the time. The model 3 qubits and up to 5 CZ reaches $\simeq37\%$ after 24h of training on a single GPU (NVIDIA V100 16GB), and $\simeq75\%$ if considering the templates with top 5 probability, which can likely improve with longer training. If we use this model as a template recommender, and "visit" the templates in the order suggested by the model, on average we would take $\simeq3.5$ visits until we reach the correct template. This is in contrast with the $63/2$ visits on average that we would have to make if we did an exhaustive template search.

\begin{figure}[h]
\centerline{\includegraphics[width=0.5\textwidth]{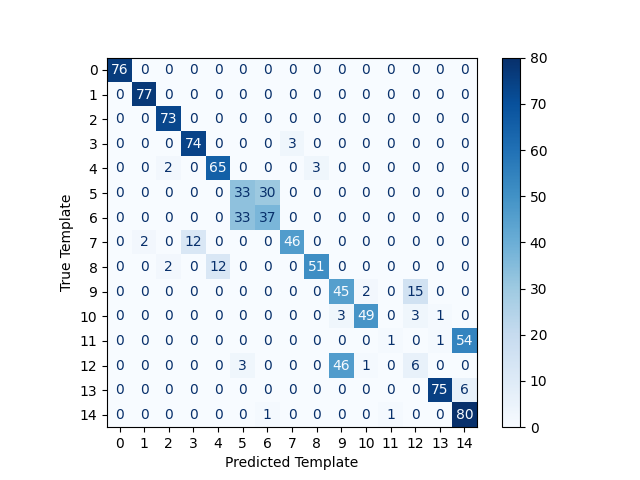}}
\caption{Confusion matrix for the 3 qubits template prediction model, for 1000 unitaries sampled uniformly from each template.}
\label{conf_matrix_3q}
\end{figure}

To understand the kind of errors the model makes, we show the confusion matrix for the 3-CZ model in Figure \ref{conf_matrix_3q}.

In the table we can see that the model guesses correctly for most of the templates, but fails for specific templates. In particular, the model usually has problems telling apart templates  with the same CZ gates but where they appear in different order, such as templates 9 and 12 shown in Figure \ref{templates_3q}. This is likely due to these templates generating unitaries that have a significant overlap, and might improve with more careful generation of the dataset. 

\begin{figure*}[htbp]
\centering
\includegraphics[width=1.0\linewidth]{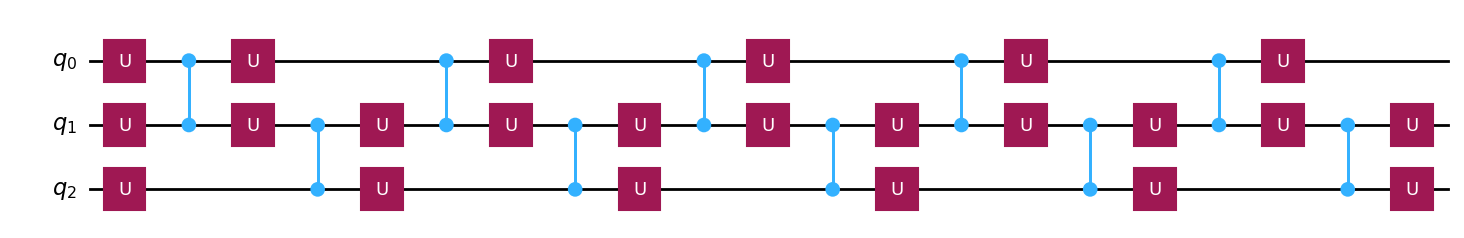}
\caption{Template used for the 3 qubit, 10 layer parameter suggestion model. Each single qubit unitary gate is decomposed as three RZ gates separated by SX gates, following the instruction set of ibm\_torino \cite{ibmQuantum_ibm_torino}.}
\label{template_3q5u}
\end{figure*}

\subsection{Parameter suggestion}

For parameter suggestion, we train models for the 4 different two-qubit templates, and for 6 and 10 layer 3-qubit templates. The 10 layer 3-qubit template is shown in Figure \ref{template_3q5u}.

For the two qubit templates, the models achieve a very high fidelity of 0.95 on average. For the three qubit templates, the models achieve around 0.5 fidelity after around 24h of training. The results are summarized in Table \ref{tab:param_results}.

\begin{table}
    \centering
    \begin{tabular}{|c|c|c|c|}
    \hline
    \textbf{\textit{\#Qubits}}&\textbf{\textit{\#Template layers}} & \textbf{\textit{\#Template parameters}}& \textbf{\textit{Avg. Fidelity}}\\
    \hline
         \multirow{4}{*}{2}&  0&  6& 0.98\\
         &  1&  12& 0.96\\
         &  2&  16& 0.96\\
         &  3&  20& 0.95\\
    \hline
         \multirow{2}{*}{3}&  6&  45& 0.52\\
         &  10&  69& 0.51\\
    \hline
    \end{tabular}
    \caption{Parameter suggestion model results}
    \label{tab:param_results}
\end{table}

\begin{figure}[htbp]
\centering
\includegraphics[width=0.9\linewidth]{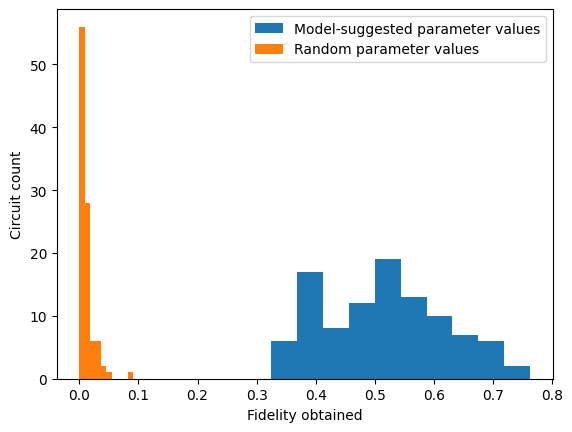}
\caption{Fidelity achieved at the starting point of the optimization by the parameter suggestion model for the 3-qubit, 10-layer template, on a set of 100 3 qubit unitaries sampled from the template. For comparison, we show the fidelity achieved on the starting point by selecting the parameters at random.}
\label{param_results_3q5}
\end{figure}

In Figure \ref{param_results_3q5} we show the fidelities corresponding to the parameters suggested by the model for the 10 layer 3-qubit template, for 100 random unitaries, compared against the fidelities obtained by selecting random parameters. Although the results may still improve with further training, we see that the suggestions are already useful when paired with the optimization procedure to refine the results.

\begin{figure}[htbp]
\centering
\includegraphics[width=0.9\linewidth]{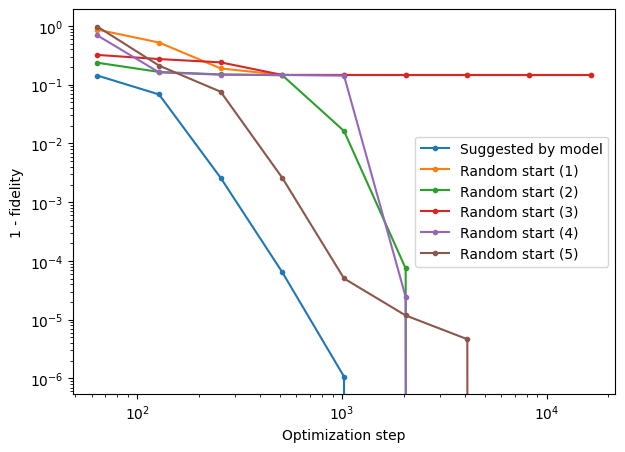}
\caption{Unitary reconstruction error as a function of the gradient descent optimization step for the Toffoli gate, using the parameters suggested by the model vs using 5 different random starting points.}
\label{toffoli_3q5}
\end{figure}

Figure \ref{toffoli_3q5} shows the result of compiling the Toffoli gate for the same template by using different starting points. For the parameters selected by the model, we see that the fidelity starts high and quickly converges to 1 at around 1000 iterations. For comparison, out of the 5 random starting points shown, two got stuck in a local minimum and the other three converged to fidelity 1, but took ~20-30\% more iterations.

\section{Discussion}

In this paper we have described an AI-based procedure for approximate synthesis of unitaries. We address the problem in three stages: template selection, parameter suggestion, and parameter refinement, and propose AI-based methods for the first two stages.

We demonstrate the idea by training two and three qubit models for the two stages. The results look in general promising: for the two qubit models, we reach almost perfect accuracy for template selection, and very high fidelity for the parameter suggestion. The accuracy and fidelity results for 3 qubits are not as high, but the models we show have only been partially trained for 24h (on a single GPU), and as described in Section III they already provide advantage over exhaustive template search and random parameter initialization. 

Interestingly, we have not found the model capacity to be a limited factor in the learning, at least in the range we have tried, and we have been able to train relatively small models (less than 1MB) that still produce high accuracy. This hints at the possibility that further improvements can be made by a more careful generation of the datasets and by tuning the training hyperparameters.

\section*{Acknowledgments}
We would like to thank Ali Javadi for the useful discussions on the topic of approximate compiling of unitaries.

\bibliographystyle{IEEEtran}
\bibliography{references}
\end{document}